\documentclass[aps,pra,twocolumn,superscriptaddress,floatfix]{revtex4}
\usepackage{epsfig,color,graphicx}
\usepackage{amsmath,amssymb,amsfonts,amsthm}
\begin{document}

\vspace{2.cm}
\title{Ground-state reference systems for expanding correlated fermions in one dimension}
\author{F. Heidrich-Meisner} 
\affiliation{Institut f\"ur Theoretische Physik C, RWTH Aachen University, 52056 Aachen, Germany}
\affiliation{Materials  Science and Technology Division, Oak Ridge National Laboratory,
 Oak Ridge, Tennessee 37831, USA and\\
 Department of Physics and Astronomy, University of Tennessee, Knoxville,
 Tennessee 37996, USA}
\author{M. Rigol}
\affiliation{Department of Physics, University of California, Santa Cruz, California 95064, USA}
\affiliation{Department of Physics, Georgetown University, Washington, District of Columbia 20057, USA}
\author{A. Muramatsu}
\affiliation{Institut f\"ur Theoretische Physik III, Universit\"at Stuttgart, 70550 Stuttgart, Germany}
\author{A. E. Feiguin} 
\affiliation{Microsoft Project Q, University of California, Santa Barbara, California 93106, USA}
\author{E. Dagotto}
\affiliation{Materials  Science and Technology Division, Oak Ridge National Laboratory,
 Oak Ridge, Tennessee 37831, USA and\\
 Department of Physics and Astronomy, University of Tennessee, Knoxville,
 Tennessee 37996, USA}

\date{May 19, 2008}

\begin{abstract}
We study the sudden expansion of strongly correlated fermions in a one-dimensional lattice, 
utilizing the time-dependent density-matrix renormalization group method.
Our focus is on the behavior of experimental observables such as the density, the momentum 
distribution function, and the density and spin structure factors. As our main result, 
we show that correlations in the transient regime can be accurately described by 
{\it equilibrium} reference systems. In addition, we find that the expansion
from a Mott insulator produces distinctive peaks in the momentum distribution 
function at $k\approx \pm \pi/2$, accompanied by the onset of power-law correlations.
\end{abstract}

\maketitle
\section{Introduction}
\label{sec:intro}
The nonequilibrium properties of strongly correlated electron systems are a 
challenging subject in need of a better understanding. While experimental 
studies in this area face difficulties in the solid state context, ultracold quantum 
gases provide a controlled way to address this difficult issue. For this reason,
recent experiments employing out-of-equilibrium cold atom 
gases in optical lattices, which allows the realization of model Hamiltonians for 
strongly correlated particles (for a review, see, e.g., Ref.~\cite{reviews}), have attracted considerable 
attention \cite{kinoshita06,greiner02,transport}.

Among the  fundamental questions recently addressed in these 
experiments is the issue of thermalization in isolated quantum systems
\cite{kinoshita06,rigol07,cazalilla06,calabrese06,calabrese07,kollath07,eisler,manmana07,cramer08,eckstein08,barthel08}. In the transient regime, 
quantum quenches 
have been shown to induce a collapse and revival of coherence properties 
\cite{greiner02}, and transport measurements in different lattice systems 
have unveiled the intriguing consequences of strong correlations \cite{transport}.
The important effects of interactions
have been  observed in the expansion of bosons in one-dimensional 
(1D) lattices as well \cite{emergence,fermionization,impurity,disorder,gangardt08}. 
In the expansion from a Mott insulator (MI) state, quasi-condensates at finite momenta 
emerge \cite{emergence}, while in the hard-core regime, 
the expansion from a superfluid state leads to the dynamical fermionization of the 
bosonic momentum distribution function (MDF) \cite{fermionization}.  The latter is a generic feature of the expansion 
of harmonically trapped hard-core bosons in the absence of a lattice 
\cite{minguzzi05}. In addition, it has been shown in Ref.~\cite{buljan08} 
that, independently of the initial interaction strength, a freely expanding 
Lieb-Liniger gas always enters a strongly correlated (hard-core like) regime. 
The expansion dynamics of strongly correlated  fermions, which due 
to the spin-degree of freedom is expected to be richer,  has not yet been 
addressed, and it is the objective of this work. 

Concretely, we study the expansion of two-component interacting 
fermions in a 1D lattice. The ground state physics of these systems is
characterized by a Tomonaga-Luttinger (TL) state with power-law decaying 
correlations at any incommensurate filling. At half-filling, a 
charge gap opens and the system exhibits quasi long-range antiferromagnetic 
correlations \cite{giamarchi}. Here, we wish to elucidate  how the initial 
state of the system, being either MI or TL, affects the expansion process.
Identifying distinctive features for the MI  is of much interest 
to experimentalists in the search for the fermionic MI state.
However, our main objective is to shed light on the relation, if any, between 
these out-of-equilibrium systems and their equilibrium counterparts.
As the main result of this work, we provide evidence that 
correlations measured in nonequilibrium are quantitatively  described by 
appropriately chosen equilibrium reference systems.

The outline of the paper is the following. First, we describe the model and the numerical
procedure in Sec.\ \ref{sec:model}. Section \ref{sec:mdf} contains our results 
on the time-evolution of density profiles and the momentum distribution function for 
both TL and MI initial states. In Sec.\ \ref{sec:gs}, we investigate what the possible relation
to equilibrium system is, and we  present a comparative analysis of spin and charge correlation functions.
We also comment on the validity of our findings in other models, such as the Hubbard chain
with a nearest-neighbor repulsion, which renders the model nonintegrable.
We conclude with  a summary of our results contained in Sec.\ \ref{sec:sum}.

\section{Model and numerical method}
\label{sec:model}

The  nonequilibrium dynamics is analyzed using the adaptive 
time-dependent density-matrix renormalization group method 
(tDMRG) \cite{white04}. We consider the 1D Hubbard model with 
nearest-neighbor hopping $t$ and an on-site Coulomb repulsion $U$:
\begin{equation}
H_0 = -t\sum_{l=1}^{N-1} ( c_{l+1,\sigma}^{\dagger}  c_{l,\sigma}^{ } 
+ \mathrm{h.c.} ) + U \sum_{l=1}^{N}
n_{l,\uparrow} n_{l,\downarrow} \,. \label{eq:ham}
\end{equation}
$c_{l,\sigma}^{\dagger}$($c_{l,\sigma}$) is a fermion creation(annihilation)
operator acting on site $l$, with (pseudo-)spin index $\sigma=\uparrow,\downarrow$,  
$n_{l,\sigma}=c_{l,\sigma}^{\dagger}c_{l,\sigma}$ is the corresponding density
operator, and we define $n_l=\sum_\sigma n_{l,\sigma}$.
$N$ denotes the number of sites, $a$ is the lattice constant,
and open boundary conditions are imposed. 
We prepare an initial state with a filling $n_{\mathrm{init}}$ that is non-vanishing 
in only a portion of the system by applying a confining box-potential 
$H_{\mathrm{conf}}=\sum_{l=1}^{N} \epsilon_l n_l$. Hence, we have
$H= H_0+H_{\mathrm{conf}}$, with  $\epsilon_l=10^{6}t$ for
$l\not\in \lbrack l_0, l_1 \rbrack$ and $\epsilon_l=0$ otherwise. At time $\tau=0$, 
we turn off $H_{\mathrm{conf}}$.  $\tau$ is given in units of $1/t$, we set $\hbar$ to unity.
 
In our tDMRG runs we use a third-order Trotter-Suzuki time-evolution scheme with a time 
step of $\Delta\tau=0.005$. The discarded weight during the time-evolution is kept 
below $10^{-8}$. To simulate the longest time scales possible 
on a given system size before the
particles are reflected at the boundaries, we select an asymmetric set-up and, hence,
particles can only expand into one direction. We have checked that the same overall 
picture is observed in symmetric set-ups (see also Ref.~\cite{emergence}).

\begin{figure}[t]
\includegraphics[width=0.45\textwidth]{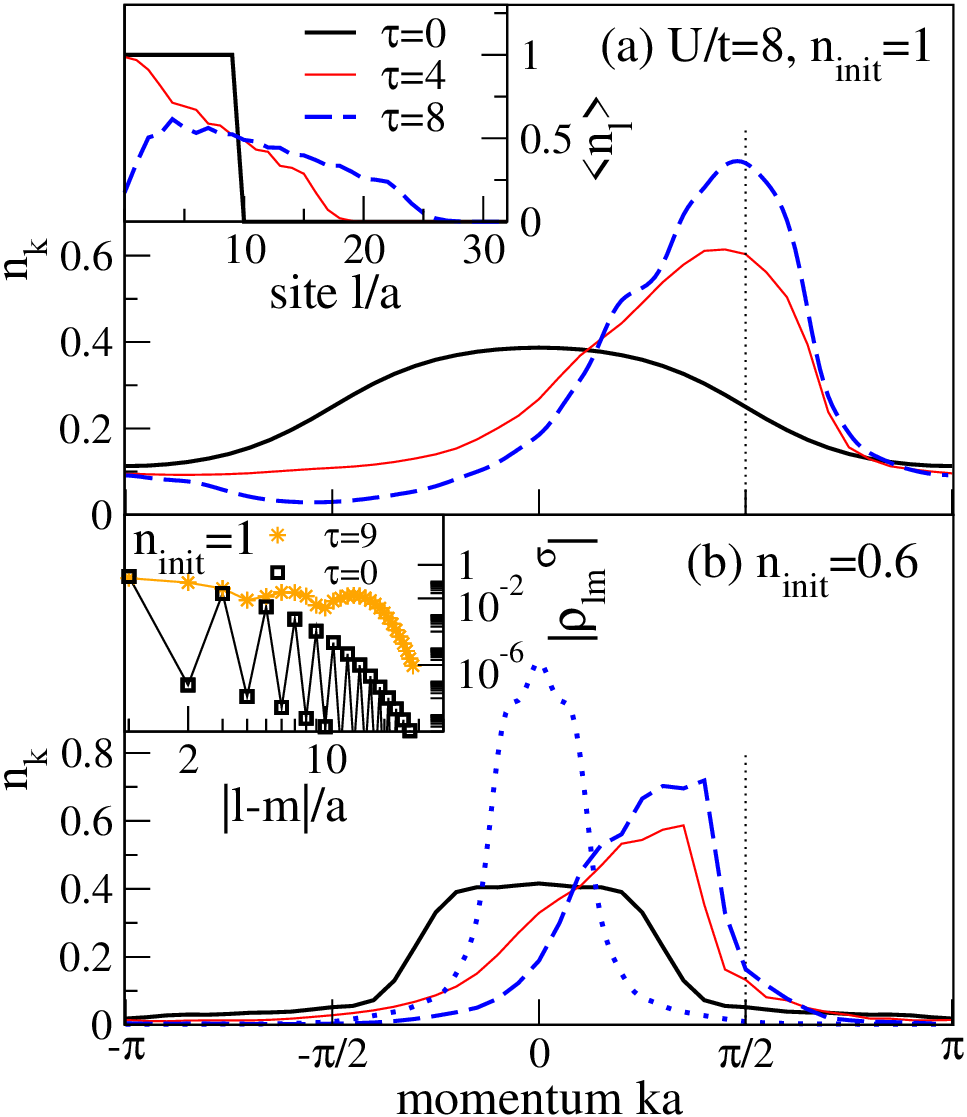}
\caption{Expansion from a:  (a) MI ($n_{\mathrm{init}}$=$1$; main panel: MDF; inset: density);
(b)  TL ($n_{\mathrm{init}}$=$0.6$); both at $U=8t$ and plotted at  times $\tau=0,4,8$. 
Note that at $\tau\gtrsim 8$ and before the right boundary is reached, 
the MDF exhibits only small changes. 
Dotted line in (b): MDF of a reference system (see text in Sec.~\ref{sec:gs}~A for details) with  $\langle n_l\rangle_{\mathrm{ref},\tau} =\langle n_l(\tau=8)\rangle$. Inset in (b): 
Decay of one-particle correlations during the expansion of the MI. The $\tau$=$0$ 
curve is for a $N$=$50$ system at half-filling. Dotted 
vertical lines in (a) and (b) denote $k=\pi/2$.}
\label{fig:nk_fermi}
\end{figure}

\section{The momentum distribution function}
\label{sec:mdf}

We first discuss the properties of the MDF $n_k$, computed from 
$n_k=(1/N) \sum_{l,m,\sigma} \exp\lbrack-ik(l-m)\rbrack \rho^\sigma_{lm}$, where 
$\rho^\sigma_{lm}=\langle c^{\dagger}_{l,\sigma} c^{}_{m,\sigma}\rangle$ is the one-particle 
density matrix. In Fig.~\ref{fig:nk_fermi}(a), we show the evolution of $n_k$ (main panel) 
and the density $\langle n_l \rangle$ (inset) for an initial MI state. The  main panel reveals 
a peculiar behavior of $n_k$: as the Mott 
insulator melts, a peak develops at a finite momentum $k_p$. 
We further find that, for $U$ larger than the band-width $W=4t$, 
$k_p$ closely approaches $\pi/2$.
This behavior resembles that of hard- and soft-core bosons 
\cite{emergence}. Qualitatively, we understand this in terms of an energy argument: in the MI 
state with $U\gg W$, the total kinetic energy is close to zero. 
Hence, particles emitted into an empty lattice have a small average kinetic energy corresponding 
to a momentum $\pi/2$. As the Fermi statistics prohibits quasi-condensation into a single 
momentum state, $n_k$ becomes a broad function around $k_p\approx \pi/2$.

While the initial MI state is characterized by an exponential decay of one-particle correlations, 
{\it i.e.}, $|\rho_{lm}^{\sigma}|$$\sim$$\exp(- |l-m|/\xi)$, $\xi$=const, we find that during the 
expansion, the system develops power-law correlations. In the inset of Fig.~\ref{fig:nk_fermi}(b),
we compare the $|\rho_{lm}^{\sigma}|$ of a MI in equilibrium with the correlations 
that emerge during its expansion, measured within the moving cloud. The inset reveals the weak 
decay of correlations during the expansion, consistent  with a power law. One may associate 
the dynamical emergence of this power law with a {\it metallization} of the moving cloud, 
which, after the melting of the MI, starts behaving as an inhomogeneous metal. 
As of now, our numerical analysis is restricted to a small number of particles and time scales of $\tau \sim 15$ only, which 
prevents us from extracting, {\it e.g.}, exponents of the power laws. Note, though, that in the case of free
fermions expanding from an insulating state with $n_{\mathrm{init}}=1$, {\it i.e.}, a state with no off-diagonal correlations, the emergence of power laws has been established for 
a large number of particles and hence over substantially larger distances than in the present work \cite{rigol05}.

\begin{figure*}[ht]
\includegraphics[width=0.42\textwidth]{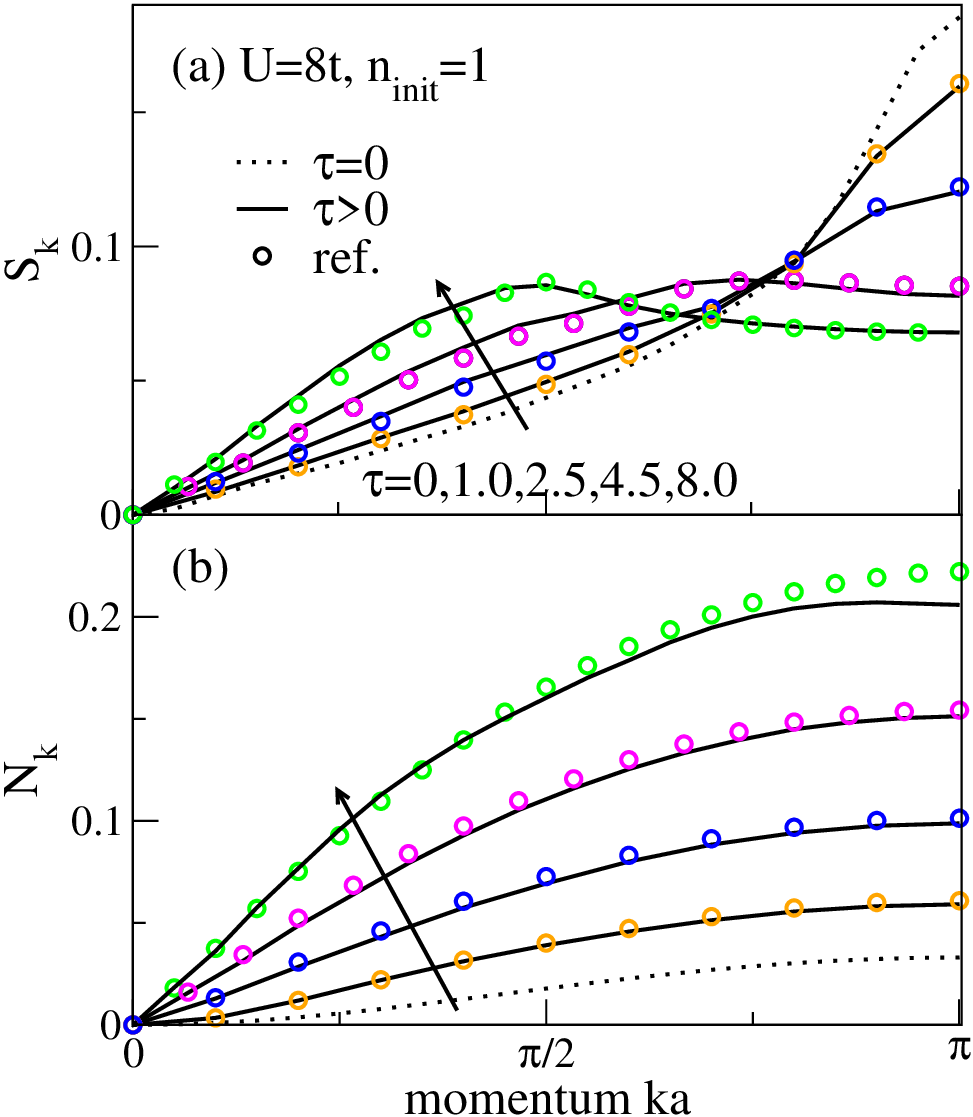}\hspace{0.5cm}
\includegraphics[width=0.42\textwidth]{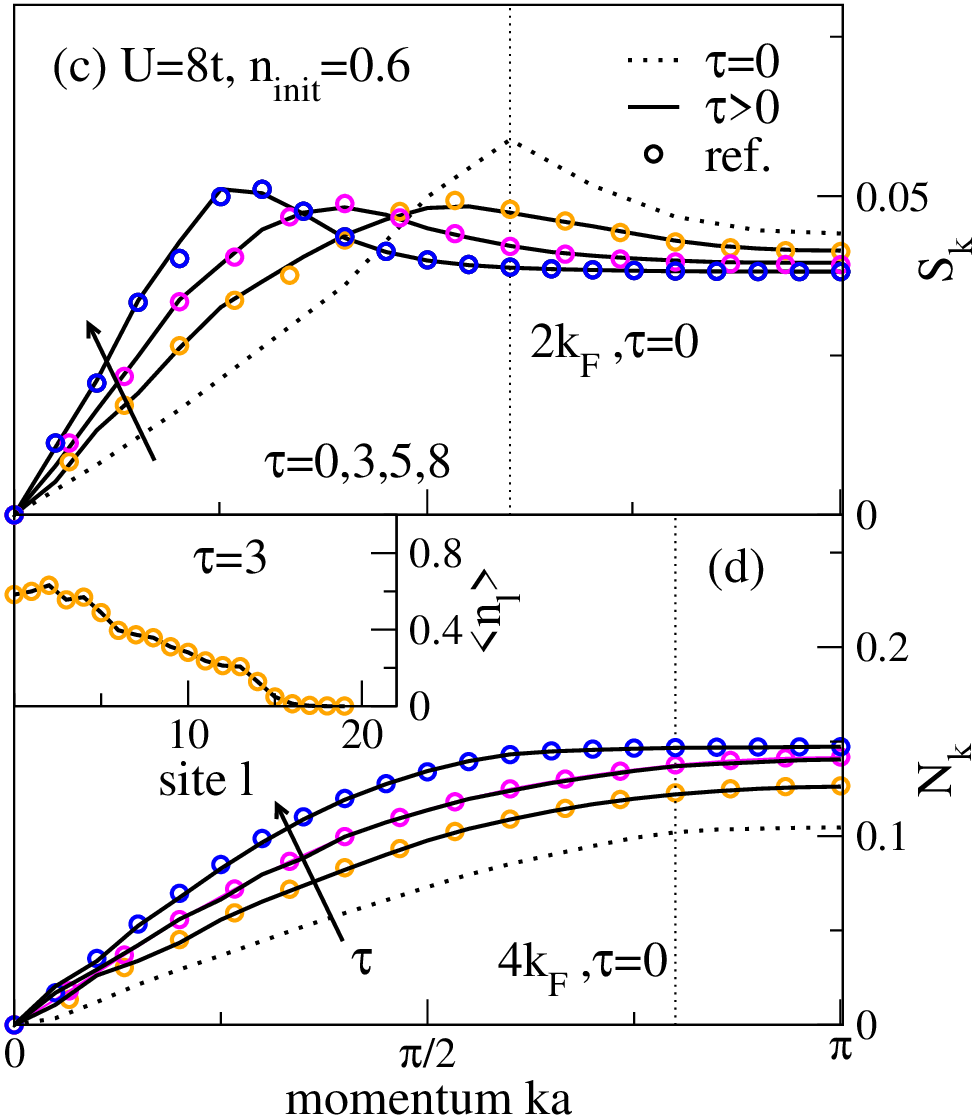}
\caption{(Color online)
Time evolution of (a) the kinetic energy $T_{\mathrm{kin}}$
and (b) the potential energy $E_{\mathrm{init}}$
for the expansion from an initial TL state with $U=8t$, $n_{\mathrm{init}}=0.6$ (dashed lines)
and the corresponding reference systems (squares).
Panel (a) further shows the kinetic energy of the cloud and of the reference systems as estimated from Eq.~(5) (thick dotted line and stars,
respectively)
as well as our approximate results for the kinetic energy $T_{\mathrm{kin}}^{\mathrm{Gal}}$ associated with the center-of-mass motion of the cloud
[solid line in (a)].
See the text in Sec.~IV C for definitions and details.}
\label{fig:n1}
\end{figure*}

In the main panel of Fig.~\ref{fig:nk_fermi}(b), we show the evolution of the MDF starting 
from a TL state with $n_{\mathrm{init}}<1$. In this case, the initial state has a well-defined Fermi momentum and a power-law decay of correlations \cite{giamarchi}. Such decay is preserved 
during the expansion. Moreover, $n_k$ also exhibits a peak, but 
at a momentum $k_p <\pi/2$ ($k_p$ increases as $n_{\mathrm{init}}\to 1$). 
Another property of this peak, distinguishing it from the peak 
formed after the melting of the MI, is that it exhibits a much sharper edge at the 
large momentum side, reminiscent of a Fermi edge.

From the previous analysis, we conclude that if 
$n_k$ could be experimentally studied during the expansion in the strongly correlated regime, 
then the emergence of peaks at $k=\pm \pi/2$ in the fermionic MDF  would serve to identify 
the presence of a Mott insulator in the initial state. The experimental challenge is to
independently control the trapping potential and the lattice \cite{kinoshita04,clement05,fort05}.
This has been achieved in the experimental study of disordered ultracold Bose gases 
in both 1D optical lattices \cite{clement05} and homogeneous 1D systems \cite{fort05}. 

At this point we would like to emphasize that the physics of our expanding system is different from the one found in theoretical studies of strongly correlated systems in 1D lattices undergoing
a  relaxation following a quantum quench \cite{rigol07,kollath07,manmana07}. In the latter, correlations have been found to decay faster than with a power law (sometimes clearly exponentially)
\cite{calabrese06,rigol07,kollath07,manmana07}, while in our moving clouds we find power-law decaying correlations [inset of Fig.~\ref{fig:nk_fermi}(b)]. 
In addition, after the relaxation to a steady state, one can ask the question of what statistical ensemble may best describe physical observables, 
but here we are solely concerned with the transient regime, {\it i.e.}, a regime in which statistical ensembles do not provide us with insights into the behavior of physical observables.

\section{Ground-state reference systems}
\label{sec:gs}

\subsection{Construction of reference systems}
\label{sec:ref}
Our results thus far have singled out a noticeable property of these systems during their
expansion: independently of the initial ground state, power laws are observed in
the nonequilibrium dynamics. In 1D systems in equilibrium,  power-law 
correlations are only seen in the ground state, as finite temperatures introduce a 
cut-off at large distances, followed by an exponential decay \cite{giamarchi}. 
Hence, one may wonder whether a system out of equilibrium can in some way resemble
the ground state of a system in equilibrium. A natural choice for such a reference 
state is the ground state of a 
system that has exactly the same density distribution as the time evolving state, 
{\it i.e.}, 
\begin{equation}
\langle n_{l}\rangle_{\textrm{ref},\tau}=\langle n_l(\tau)\rangle\,,
\end{equation}
 for all sites $l$. Hence,
a reference system has to be determined at each given time.
We  construct such reference states by self-consistently computing  a set of onsite energies 
$\epsilon_l$  in
\begin{equation}
 H_{\mathrm{ref}}= H_0 + \sum_{l=1}^N \epsilon_l n_l
\end{equation}
with $H_0$ from  Eq.~(\ref{eq:ham}) 
such that at a desired time, the density profile $\langle n_l(\tau)\rangle$ is
reproduced, while keeping $t$ and $U$ fixed. Once the density 
has converged within an error of 
\begin{equation}
\delta n_{\tau}=\sum_l |\langle n_l(\tau)\rangle-\langle
n_{l}\rangle_{\textrm{ref},\tau}|/\sum_l \langle n_l(\tau)\rangle < 10^{-3}\,,
\end{equation} 
we compare quantities of interest in both systems.

\subsection{Spin and charge structure factor}

We now turn to the comparative analysis of correlation functions. We compute the spin-spin ($S_k$) and density-density ($N_k$) structure factors, 
which are the Fourier transforms of the spin-spin ($S_{lm}=\langle S_l^z S_m^z\rangle$)
and density-density ($N_{lm}=\langle {n}_l {n}_m\rangle 
-\langle {n}_l\rangle \langle {n}_m\rangle$) correlations, respectively. 
The spin operator is defined as $S_l^z=(n_{l,\uparrow}-n_{l,\downarrow})/2$.
In equilibrium and for a homogeneous system, $S_k$
peaks at $2k_F$ while $N_k$ exhibits a kink at $4k_F$ 
\cite{giamarchi}, where $k_F=\pi n/2$ is the Fermi momentum. Consistently, 
for the two cases $n_{\mathrm{init}}=1$ and $n_{\mathrm{init}}=0.6$ shown 
in Figs.~\ref{fig:n1}(a) and \ref{fig:n1}(c), respectively, $S_k(\tau=0)$ [dotted lines]
peaks at $k=\pi$ and $k=\pi\, 0.6$, while  $N_k$ in the case of $n_{\mathrm{init}}=0.6$
has a weak kink at $k=2\pi\, 0.6 $ [dotted line in Fig.~\ref{fig:n1}(d)].
During the expansion, the peak in $S_k$ shifts to smaller momenta and the 
maximum is less sharp, as shown in Figs.~\ref{fig:n1}(a) and \ref{fig:n1}(c) 
[solid lines]. Qualitatively, we understand this behavior in terms of the decrease 
of the average density during the expansion into the initially empty lattice, 
giving rise to a shift of the $2k_F$ peak in $S_k$ and a broadening due to the inhomogeneity. 
Further, we propose an operational definition of a Fermi-momentum 
$k_F^{\tau}$ in the expanding clouds by taking the position of the peak in 
$S_k$, yielding $2k_F^{\tau}$. This supports the use of the term 
``metallization'' for the process that  fermions escaping from a 
MI undergo.

The density correlation $N_k$ does not show any particular features during the 
time-evolution, as the kink is washed out due to the inhomogeneity. $N_k$ increases 
monotonously with time, reflecting an increase of the overall charge fluctuations, 
due to the closing of the charge gap as the MI melts.

Our most remarkable finding, and thus the key result of this work, is the excellent 
agreement seen in Fig.~\ref{fig:n1} for $S_k$ and $N_k$ between the expanding 
cloud -- a genuine nonequilibrium situation -- and the inhomogeneous reference systems,  
which are in their ground state (circles in Fig.~\ref{fig:n1}). We are therefore 
led to conclude that, during the expansion, spin and charge correlations out of 
equilibrium are, to a very good approximation, the same functionals 
of the density as the ones in equilibrium systems in the ground state.  

\subsection{Time-evolution of kinetic and potential energy}
\label{sec:erg}

Intuitively, we expect that our way of preparing the reference systems must yield 
properties similar to those of the moving clouds on short time scales. However, the 
agreement between the clouds and the equilibrium systems exists both at short and long times, 
and is thus preserved during the expansion into the empty lattice. A noticeable difference 
exists between $n_k$ of the moving clouds and $n_k$ of the corresponding
reference systems. As discussed before, $n_k$ of the expanding 
cloud has a finite-momentum maximum, which is also present in the symmetric expansion,
while $n_k$ of the reference system in its ground state is symmetric around $k=0$
[see dotted line in Fig.\ \ref{fig:nk_fermi}(b)]. 
Hence, observables related to $n_k$ cannot be accounted for 
with this procedure.

Our previous findings on the spin and density structure factors may seem puzzling: 
at any given time, the sum of the kinetic, $T_{\mathrm{kin}}$, and interaction 
energy $E_{\mathrm{int}}$ of the time-evolving state is much higher than the energy of the 
reference system, which, having the same density profile, is in its ground state.
This difference grows with time. In the case of $n_{\mathrm{init}}=0.6$ and $U=8t$, 
we have $E_0=T_{\mathrm{kin}}+E_{\mathrm{int}}=-6.86t$, which of course is a constant 
in time. At $\tau=8$, this splits into 
$T_{\mathrm{kin}}=-t \sum_{l,\sigma} \langle c_{l+1,\sigma}^{\dagger}c_{l,\sigma}+h.c.\rangle= -6.97t$
and 
$E_{\mathrm{int}}=U\sum_{l}\langle n_{l,\uparrow}n_{l,\downarrow}\rangle = 0.11t$.
In contrast, for the reference system, we find $T_{\mathrm{kin}}^{\mathrm{ref}}=-10.93t$,
and $E_{\mathrm{int}}^{\mathrm{ref}}=0.18t$, adding up to 
$E_0^{\mathrm{ref}}= -10.75t$. Thus, the main difference is due to the kinetic energies.

\begin{figure}[t]
\includegraphics[width=0.48\textwidth]{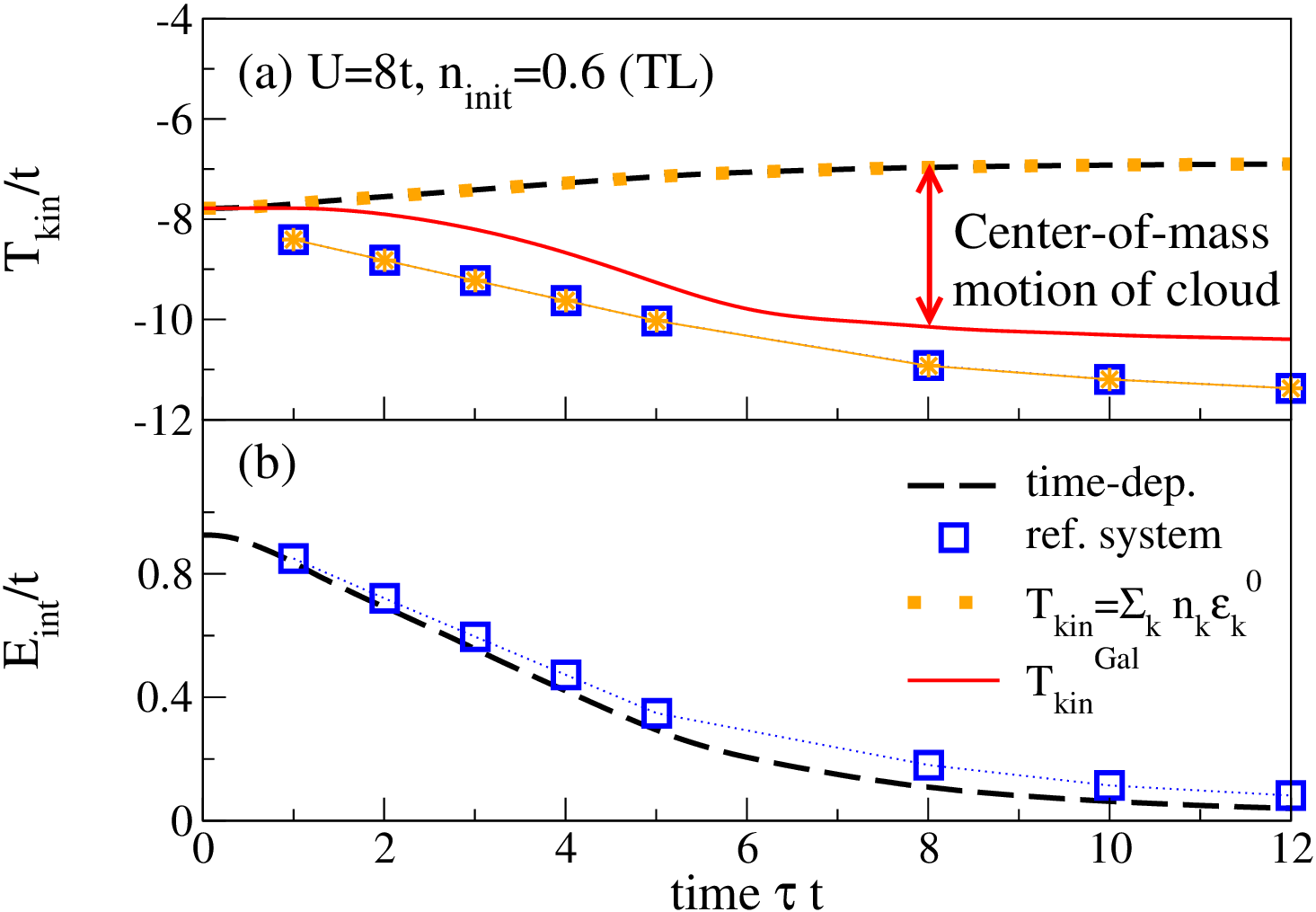}
\caption{(Color online) Time evolution of (a) the kinetic energy $T_{\mathrm{kin}}$ 
and (b) the potential energy $E_{\mathrm{init}}$ 
for the expansion from an initial TL state with $U=8t$, $n_{\mathrm{init}}=0.6$ (dashed lines)
and the corresponding reference systems (squares).
The plot further shows the kinetic energy of the cloud as estimated from Eq.~(\ref{eq:est}) [thick dotted line and stars]
as well as our approximate results for the kinetic energy $T_{\mathrm{kin}}^{\mathrm{Gal}}$ associated with the center-of-mass motion of the cloud (solid line).
See text in Sec.~\ref{sec:erg} for definitions and details.
}
\label{fig:erg}
\end{figure}

We argue that both systems can be related by a Galilean transformation and, thus, understand 
why their structure factors are similar. This means that the difference between the kinetic 
energies $T_{\mathrm{kin}}$ and $T_{\mathrm{kin}}^{\mathrm{ref}}$ is mainly due to 
the average momentum of the moving cloud [$k_0=\sum_k k n_k/\sum_k n_k $], {\it i.e.}, that $T_{\mathrm{kin}}^{\mathrm{ref}}\approx T_{\mathrm{kin}}^{\mathrm{Gal}}$, the latter being 
the kinetic energy of the particles in a reference frame moving with the cloud. 
To prove this, we first notice that, using the 
MDF, the kinetic energy can be estimated  as 
\begin{equation}
T_{\mathrm{kin}}=\sum_k n_k \epsilon_{k}^{0} \,, \label{eq:est}
\end{equation} 
where $\epsilon_k^{0}=-2t\cos k$ 
is the dispersion relation in the noninteracting case. This assumption leads to 
$T_{\mathrm{kin}}\approx -6.97t$ and $T_{\mathrm{kin}}^{\mathrm{ref}} \approx-10.93t$, 
as estimates for the kinetic energy of the expanding and reference systems at $\tau=8$
($n_{\mathrm{init}}=0.6$, $U=8t$), respectively. Both values are very close to the 
exact results presented before. We then compute 
$T_{\mathrm{kin}}^{\mathrm{Gal}}=\sum_k n_k \epsilon_{k-k_0}^{0}=-10.15t
\approx T_{\mathrm{kin}}^{\mathrm{ref}}$ at $\tau=8$, which corroborates our interpretation: 
the energy difference is mostly due to the finite momentum of the cloud, and not due to 
contributions of the internal kinetic or interaction energy. This picture is further supported
by Fig.~\ref{fig:erg} that contains the time-evolution for $T_{\mathrm{kin}}$, $E_{\mathrm{int}}$, $T_{\mathrm{kin}}^{\mathrm{ref}}$,
and $T_{\mathrm{kin}}^{\mathrm{Gal}}$ for the parameters discussed in this section.

\begin{figure}[t]
\includegraphics[width=0.48\textwidth]{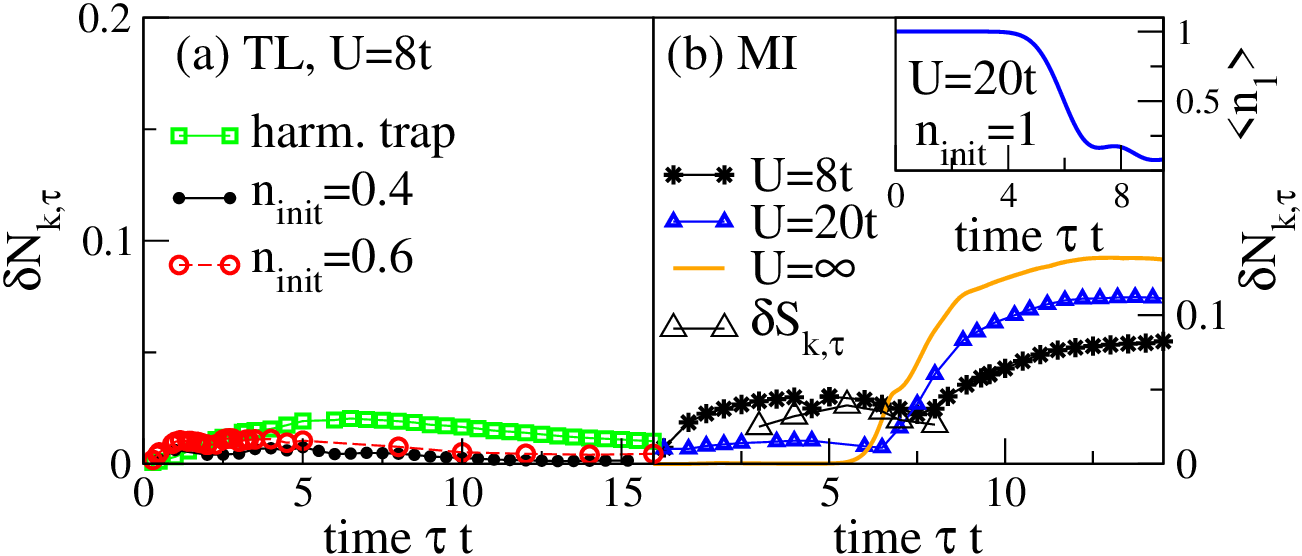}
\caption{(Color online) Relative deviation of the density structure factor vs. time [see Eq.~(\ref{eq:nk})], for:  (a)  
$n_{\mathrm{init}}$=$0.4,0.6$ and  a harmonic trap 
(8 fermions, $V_{\mathrm{trap}}$=$0.02t$, $U$=$8t$) and (b) $n_{\mathrm{init}}$=$1$ ($U/t$=$8,20,\infty$). 
We estimate our calculation of $\delta N_{k,\tau}$ to have an accuracy of $\pm 0.01$ in the worst case. Inset in (b): density in the leftmost site vs. time for $U=20t$.}
\label{fig:long}
\end{figure}

\subsection{Breakdown of the ground-state reference system description}

It is next important to identify conditions for a breakdown of the reference-system description. 
To this end, we study the relative difference between the time-dependent and the reference 
systems' density structure factors, 
\begin{equation}
\delta N_{k,\tau}= {\sum_k |N_k(\tau) - N_{k,\tau}^{\mathrm{ref}}|}/{\sum_k N_k(\tau)}.
\label{eq:nk}
\end{equation}
The corresponding errors in $S_k$ are smaller than those in $N_k$, and we thus concentrate 
on the latter. Let us start with the   initial TL. We consider two cases: first, 
$n_{\mathrm{init}}<1$ in a box trap [see Fig.~\ref{fig:n1}(d)]. Second, as such a set-up 
is more realistic to account for experiments, we follow the evolution of fermions escaping 
from a harmonic trap $V_{\mathrm{trap}}\sum_l (l)^2 n_l$. From the results displayed in 
Fig.~\ref{fig:long}(a), our key observation is that  $\delta N_{k,\tau}\lesssim 0.02$
remains very small in both cases. Hence, for an initial TL state and for both $N_k$ and $S_k$, 
the description given by the equilibrium systems is very good up to the largest times simulated.

We next turn to the case of an initial MI region, and  present results 
in Fig.~\ref{fig:long}(b) for $U=8t,20t,\infty$. The $U=\infty$ case is treated with 
exact diagonalization, after mapping the charge sector of our two-component fermion system 
to spinless fermions \cite{giamarchi}. A behavior similar to the TL case is found at 
times $\tau \lesssim 5$, with $\delta N_{k,\tau}\lesssim 0.04$. However, for times after 
the melting of the MI region [$\tau \gtrsim 5$, see the inset in Fig.~\ref{fig:long}(b)], 
a substantial increase of $\delta N_{k,\tau}$ becomes evident, as shown in Fig.\ \ref{fig:long}(b). Thus, for the 
MI expansion, reference systems work well only up to the point at which the Mott 
insulator totally melts.

\begin{figure}[t]
\includegraphics[width=0.48\textwidth]{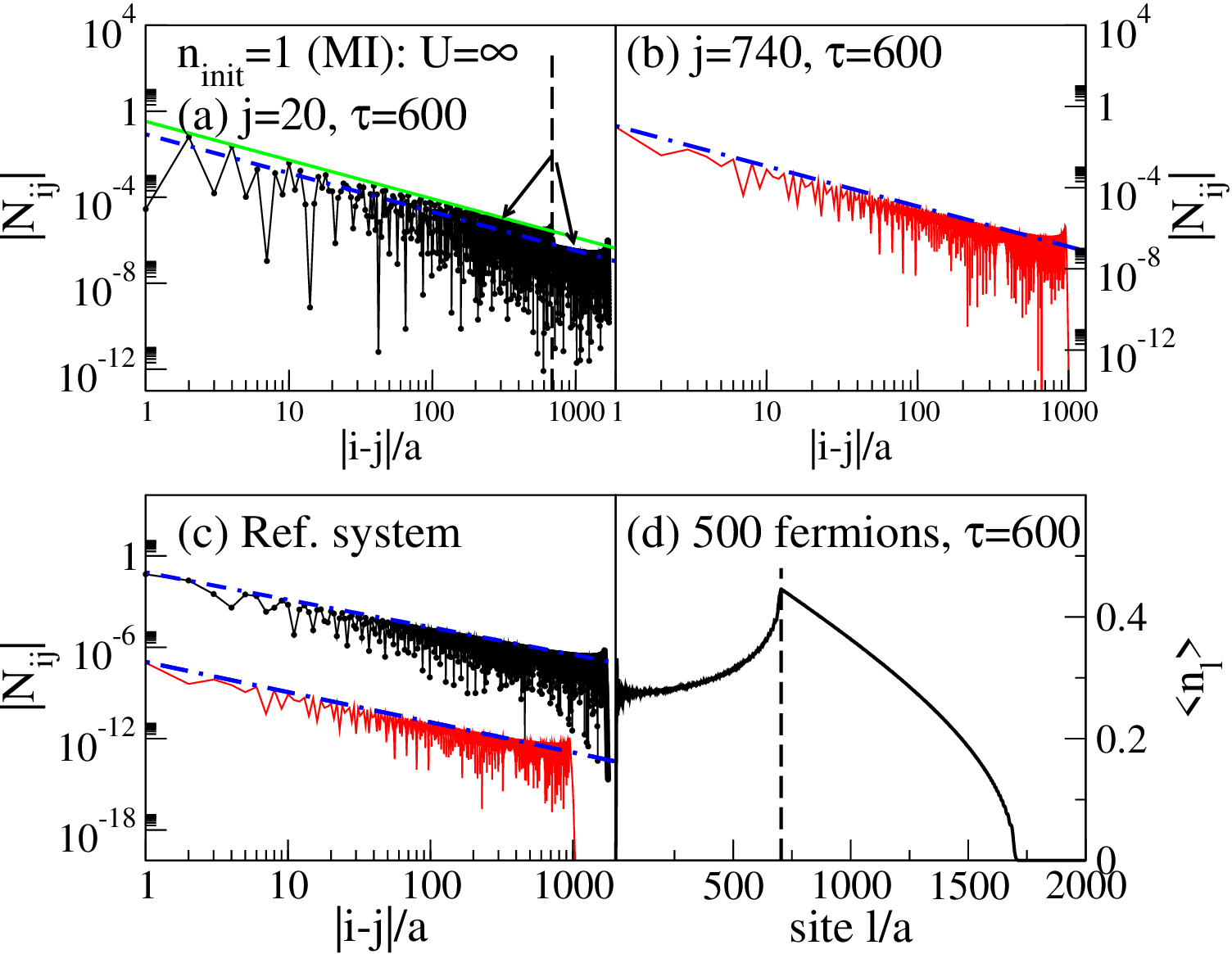}
\caption{(Color online) $U=\infty$, 500 particles, $n_{\mathrm{init}}=1$. (a) 
Decay of $|N_{ij}|$ at time $\tau=600$ for $j=20$ (solid line with squares).
The dashed lines are fits to the envelope of $|N_{ij}|$ using $f(|i-j|)=\alpha |i-j|^{-\beta}$. Note that two regions
 appear that are described by the same exponent $\beta$, but a different prefactor $\alpha$.
(b) The same as in (a), but for $j=740$ (solid line). Note that we plot correlations for $i>j$ only. (c) Density-density correlations measured in the reference system
for $\tau=600$, with $j=20$ (solid line with squares) and $j=740$ (solid lines). The latter curve has been offset for clarity.
The dashed-dotted lines are the fits from panels (a) and (b). (d) Density profile at time $\tau=600$. Vertical, dashed lines
in (a) and (d) mark $i=705$, separating the regions with different prefactors in the power-law decay of $|N_{ij}|$ at the time $\tau=600$ considered in this plot.}
\label{fig:inf}
\end{figure}

This deviation of the time-dependent data from the reference systems is associated to the 
appearance of particular coherence properties in the portion left behind by the moving 
cloud after the melting of the MI, a feature that is not captured by the reference systems. To substantiate this interpretation,
we present results for the decay of $N_{ij}$ in the $U=\infty$ limit in Fig.~\ref{fig:inf}.
In that limit, we are able to exactly treat arbitrary time scales for both large number of particles and a large system size. 
The data displayed in Fig.~\ref{fig:inf} are for 500 spinless fermions expanding from a MI region into a lattice of 2000 sites.  

In Fig.~\ref{fig:inf}, we measure the density-density correlations at a time  $\tau=600$, at which the initial Fock 
state has already completely melted. Figure~\ref{fig:inf}(d) shows the  density profile at this time
and from that plot, we see  that at time $\tau=600,$ $i\approx 705$ separates the front of fast particles from slower ones, as indicated by the vertical dashed line in the figure.
Panels (a) and (b) show $|N_{ij}|$ for $i>j$, and $j=20$ and $j=740$, respectively,  {\it i.e.},
measured from behind and from inside the moving front [Fig.~\ref{fig:inf}(d)]. While in the latter case, clearly a unique power-law decay of $|N_{ij}|$ is observed, the former case is more involved. 
There, the envelope of the correlator also decays according to $|N_{ij}|=\alpha \, |i-j|^{\beta}$, 
yet for $i\lesssim 705$ and $i\gtrsim 705$, a different prefactor $\alpha$ is found. These are  the two regions in Fig.\ \ref{fig:inf}(a) indicated
by the arrows. 
The exponent $\beta=2K$ is universal and expected to be $\beta=2$ since the Luttinger parameter $K$ of spinless fermions is $K=1$, but the prefactor  -- {\it in the ground-state} -- is
essentially a function of the average density, or the Fermi momentum, respectively \cite{giamarchi}. The density-density correlations therefore exhibit a distinctly different behavior comparing the moving
front of fast particles ($i\gtrsim 705$) and those left behind ($i\lesssim 705$).  It is exactly this step-like feature, {\it i.e.},
the sudden change in the prefactor of the power law followed by $|N_{ij}|$, 
that is not captured by the reference systems. This is revealed in Fig.~\ref{fig:inf}(c), which shows the $|N_{ij}^{\mathrm{ref}}|$ as measured in
the reference system constructed for time $\tau=600$, for both $j=20$ and $j=740$. The plot includes the fits to $|N_{ij}|$ from panels (a) and (b) [dot-dashed lines]. 
While for $j=740$, $|N_{ij}|$ is well described by the reference systems, $|N_{ij}^{\mathrm{ref}}|$ with $j=20$ does not show the step like feature observed in the moving cloud, 
as the reference systems fail to account for the separation of particles moving at different velocities.   

For this reasoning to apply, it is important to realize that such a separation of velocities as reflected in the two prefactors to the power-law decay is not
observed in the expansion  from a TL state. There, a  power-law with a single pair of exponent and prefactor governs the decay of one-particle and density-density correlations  \cite{fermionization}.

\subsection{Nonintegrable systems}
\label{sec:nonint}

Beyond the case of the Hubbard model Eq.~(\ref{eq:ham}), the question arises whether non-stationary states 
of other model Hamiltonians may as well be described by ground-state reference systems. Conceptually, one
may wonder whether integrability plays a role or not.

\begin{figure}[t]
\includegraphics[width=0.48\textwidth]{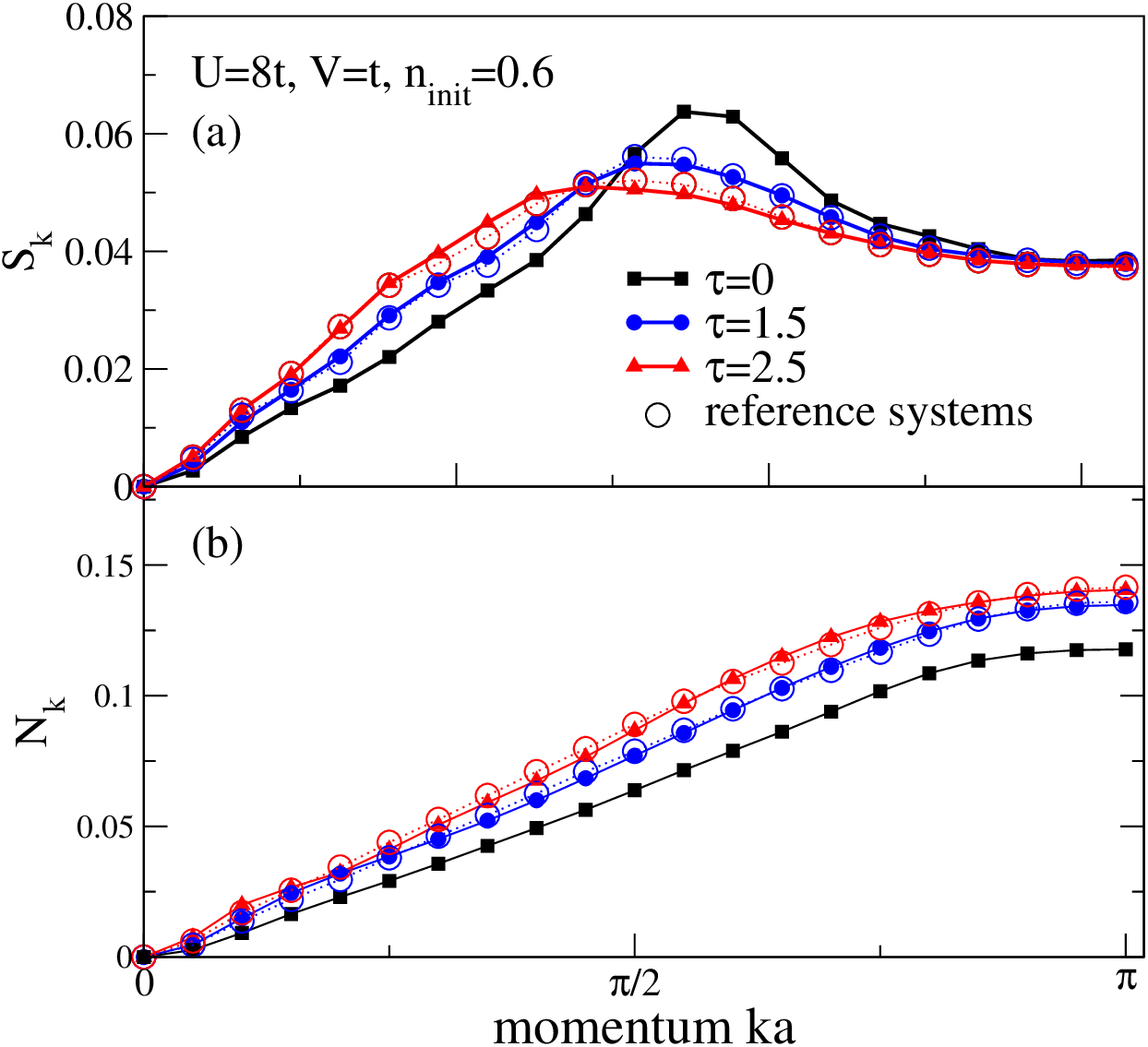}
\caption{(Color online) Structure factor for (a) spin and (b) density correlations for the extended Hubbard model with
$U=8t$, $V=t$. The results are for the expansion from an initial state with $n_{\mathrm{init}}=0.6$ and 6 particles.
Time-dependent data are represented with solid lines and solid symbols, the  corresponding ground state reference system
ones are displayed with open circles.
}
\label{fig:nonint}
\end{figure}

While a full account of these interesting issues is beyond the scope of the present work, we wish to at least comment  on one
additional model, the extended Hubbard model. In addition to the terms given in Eq.~(\ref{eq:ham}), this model incorporates a nearest-neighbor
repulsion $H_{\mathrm{2}}$:
\begin{equation}
H_{2} = V \sum_{l=1}^{N-1}  n_{l} n_{l+1} \,.\label{eq:tuv}
\end{equation}
The nearest-neighbor interaction both renders the system nonintegrable and induces additional phases at half-filling. 
For $V<U/2$, the system is a Mott insulator, while a large $V$ drives the system into a charge-density wave phase \cite{jeckelmann02}.
Here we focus on the numerically less demanding case of the expansion from an initial state with an incommensurate filling of 
$n_{\mathrm{init}}=0.6$. We postpone the discussion of the MDF to a future publication, but rather
compute the spin and charge structure factors for both the expanding system and  the reference systems constructed 
according to the prescription of Sec.~\ref{sec:ref}. The results for $U=8t$ and $V=t$ are collected in Fig.~\ref{fig:nonint}, and we note that in this case,
we consider the expansion from a box trap only.

Our observation is that the reference systems still provide a very good approximation 
to the time-dependent correlations. Moreover, the agreement remains to be 
much better for the spin structure factor than for the charge structure factor. Future work will 
have to clarify whether the reference system description breaks down as $V$ is increased. 
In conclusion, we find that the reference system description does not seem to be crucially 
dependent on integrability, and we expect a similar picture to emerge in other models.

\subsection{Discussion: Relation to density functional theory}

Our main results have shown that starting from a Mott-insulator, the temporal evolution of spin and density correlation functions are very accurately described 
by the ground-state of reference systems defined at each instant of time so that they have the same density distribution as the time evolving system. 
Such a description deteriorates after the Mott insulating region has totally melted. On the other hand, for systems starting with densities
$n_{\mathrm{init}} < 1$, the description with reference systems remains valid up to the largest times reached in our simulations. These results explicitely show that
the correlation functions studied here are functionals of the density, a fact that is in accordance with density functional theory (DFT) for time-dependent systems
\cite{dft,tdft}.

DFT considers, both for the ground state and time-dependent situations, this kind of Hamiltonian \cite{dft,tdft}:
\begin{equation}
H= H_{\mathrm{kin}} + H_{\mathrm{Coulomb}} + H_{\mathrm{ext}}\,\label{eq:dft}\,,
\end{equation}
{\it i.e.}, one separates the Hamiltonian into kinetic energy, the interaction energy due to Coulomb interactions, and an external potential.
We should here remark that in our case, the time-dependent
external potential is discontinuous at time $\tau=0$, as we use $H_{\mathrm{ext}}=H_{\mathrm{conf}}$ at time $\tau=0$ and  $H_{\mathrm{ext}}=0$  for $\tau>0$.
Therefore,
it does not strictly comply with the assumptions needed to prove the Runge-Gross theorem \cite{runge84}, {\it i.e.}, the external potential needs to be
analytical 
around $\tau=0$.

However, and most importantly,
the reference systems explicitely provide the required functionals, namely the correlation functions in the respective ground-states. This is, to our
opinion, a rather surprising and nontrivial fact that the correlations of a genuinely non-stationary state can be quantitatively described by equilibrium
systems. Moreover, since they are in their ground-state, this suggests that
a minimum principle is at work here. As shown in Sec.~\ref{sec:nonint}, these conclusions
are not restricted to the pure Hubbard-model, {\it i.e.}, they are not a consequence of integrability. 
Therefore, we expect that they hold in general.

\section{Summary}
\label{sec:sum}

In this work, we have identified several remarkable and unexpected properties of fermions expanding into an empty lattice. 
These include the emergence of coherence  as well as well as an accumulation of 
particles at momentum $\pi/2$ in the expansion of particles coming from a MI region.
In particular, 
we have shown that correlation functions of expanding, interacting fermions can be accurately described by equilibrium reference systems in their ground state. 
These results are expected to qualitatively carry over to other models as well, and certainly also apply to the case of hard-core bosons.

\begin{acknowledgments}
We thank  A.~Kolezhuk, A. Leggett, and D.~Scalapino  for fruitful discussions. 
F.H.-M. and E.D.  were supported in part by the NSF grant No.\ DMR-0706020 
and the Division of Materials Science and Engineering, U.S. DOE, under contract 
with UT-Battelle, LLC. M.R. was supported by the NSF grant No.\ DMR-0706128 and the 
Department of Energy Grant No.\ DOE-BES DE-FG02-06ER46319. 
A.M. acknowledges partial support by the DFG through SFB/TRR21.
\end{acknowledgments}

\end{document}